# The Journey to Serverless Migration: An Empirical Analysis of Intentions, Strategies, and Challenges

Muhammad Hamza[1][0000-0002-1633-9995], Muhammad Azeem Akbar[1][0000-0002-4906-6495], Kari Smolander[1][0000-0002-7043-0458]

[1] Department of Software Engineering, LUT University, Lappeenranta, Finland
Muhammad.hamza@lut.fi, Azeem.akbar@lut.fi, Kari.smolander@lut.fi

**Abstract.** Serverless is an emerging cloud computing paradigm that facilitates developers to focus solely on the application logic rather than provisioning and managing the underlying infrastructure. The inherent characteristics such as scalability, flexibility, and cost efficiency of serverless computing, attracted many companies to migrate their legacy applications toward this paradigm. However, the stateless nature of serverless requires careful migration planning, consideration of its subsequent implications, and potential challenges. To this end, this study investigates the intentions, strategies, and technical and organizational challenges while migrating to a serverless architecture. We investigated the migration processes of 11 systems across diverse domains by conducting 15 in-depth interviews with professionals from 11 organizations. we also presented a detailed discussion of each migration case. Our findings reveal that large enterprises primarily migrate to enhance scalability and operational efficiency, while smaller organizations intend to reduce the cost. Furthermore, organizations use a domain-driven design approach to identify the use case and gradually migrate to serverless using a strangler pattern. However, migration encounters technical challenges i.e., testing event-driven architecture, integrating with the legacy system, lack of standardization, and organizational challenges i.e., mindset change and hiring skilled serverless developers as a prominent. The findings of this study provide a comprehensive understanding that can guide future implementations and advancements in the context of serverless migration.

## 1 Introduction

Serverless has emerged as a disruptive cloud computing paradigm that revolutionized the way software applications are developed and deployed. Companies utilizing a serverless paradigm can solely focus on developing the application logic rather than maintaining and provisioning the underlying cloud infrastructure [1]. Function-as-a-service (FaaS), an implementation serverless pattern, enables developers to create an application function in the cloud that automatically triggers in response to an event or request [1]. Companies in serverless computing are only charged for the resources their applications consume. This pricing model contrasts traditional cloud computing, where resources are persistently leased, irrespective of whether the application is running. Seeking the potential of this computing paradigm, major cloud providers such as AWS Lambda, Google Cloud Functions, Microsoft Azure have rolled out their serverless platforms with well-defined features and pricing.

According to recent reports, the serverless market will substantially grow from $3 billion in 2017 to an approximate value of $22 billion by 2025 [2]. Moreover, it is estimated that 50% of enterprises worldwide are expected to adopt serverless computing by 2025 [3]. However, the distinct features of serverless computing also bring new challenges (e.g., statelessness, short-lived functions) for companies developing greenfield projects or migrating existing legacy applications. Legacy applications face a myriad of technical challenges (e.g., excessive coupling, maintenance complexities) and business-related challenges (e.g., longer release time, low developers' productivity). Therefore, embracing a migration to serverless computing offers a compelling solution to address current challenges while improving system scalability and faster releases to the market.

Significant research has been done on the various aspects of serverless computing, including serverless architectural design [4], development features, technological aspects, performance characteristics of serverless platforms [5], etc. For instance, Lin et al. [6] extensively discussed serverless architecture, presenting its associated concepts, advantages, and disadvantages while exploring various architectural implications. Taibi et al. [7] conducted a thorough multivocal literature review to identify a set of patterns (common solutions to solve common problems) and classified them based on the benefits and issues. Similarly, Wen et al. [8] conducted a systematic literature review and highlighted the benefits of serverless computing, its performance optimization, commonly used platforms, research trends, and promising opportunities in the field. Another study by Taibi et al. [9] discussed the increasing popularity and adoption of serverless architecture, its evolution, benefits, and challenges associated with this new cloud computing paradigm. However, to the best of our knowledge, no study empirically investigates why companies migrate to serverless architecture, how they refactor their legacy applications, and what technical and organizational challenges they face during migration to serverless.

To this end, we conducted 15 in-depth interviews with industrial practitioners from 11 organizations in different countries. All the practitioners were involved in the migration or development of serverless applications. Thus, the main objective of this study is to uncover the intentions of migrating applications to serverless, refactoring strategies, and technical and organizational challenges. Furthermore, we talked about 11 serverless-based systems, two being during migration, eight having migrated over the past two years and one greenfield development. The research question addressed in this study are:

**RQ1:** What are the intentions of migrating legacy applications to serverless architecture?

**RQ2:** What are the migration strategies that companies employ?

**RQ3:** What technical and organizational challenges do companies face during migration to serverless architecture?

The rest of paper is structured as follows: Section 2 covers related serverless work; Section 3 outlines our research method; Section 4 details the interview cases; Section 5 discusses results, and Section 6 concludes the study.

## 2      Related Work

Serverless computing represents a paradigm shift in the way applications are developed and deployed, eliminating the need to manage underlying infrastructure. The unique nature of this architecture has led researchers and practitioners to explore its facets, seeking to optimize it for future use. Existing studies have discussed different aspects of serverless computing, including architectural design, performance improvement, technological aspects, testing and debugging [10] [11], and empirical investigations.

For instance, Wen et al. [12] mined and manually analyzed 619 discussions from the stack overflow. The study finds the challenges that developers face when developing the serverless application. Our study is methodologically different from this as we analyzed the migration process of 11 systems by conducting 15 interviews from 11 organizations. We believe that developers do not fully share their migration experiences on these Q&A sites; instead, they share their problems and get feedback from community members.

Similarly, Eskandani and Salvaneschi [13] provided insight into the FaaS ecosystem by analyzing the 2k real-world open-source applications developed using a serverless platform. The study collected open-source applications from GitHub. The study explores aspects like the growth rate of serverless architecture, architectural design, and common use cases. Another similar study by Esimann et al. [14] revealed various aspects, including implementation, architecture, traffic patterns, and usage scenarios. The study is slightly different from Eskandani and Salvaneschi [13]. Both studies analyze the serverless open-source applications developed by the greenfield approach. These studies did not discuss the migration process to serverless, intentions, and technical and organizational challenges.

Taibi et al. [15] identified the bad practices among the practitioners while developing serverless applications and proposed a solution to those practices. They identified seven concerns, derived five bad practices, and proposed solutions to address them. Our work is completely different from this study as we investigated the migration process of legacy applications to serverless. Leitner et al. [16] conducted a mixed-method empirical study highlighting the need for a different mental model when adopting serverless and prevalent application patterns. However, the study did not investigate the migration processes. Moreover, the study was conducted in 2018, and since then serverless computing has evolved rapidly, and many companies have migrated their systems between 2019 to date [2]. To the best of our knowledge, no existing study investigated the serverless migration processes. This study explores migration processes by investigating the intention of migration, strategies, and challenges (technical and organizational) when migrating legacy applications to serverless.

## 3 Research Method

We employed a qualitative research method, specifically semi-structured interviews, to fulfill the objective of this study. Qualitative approaches aim to understand real-world situations, deal directly with complex issues, and are useful in answering "how" questions in the study [17]. The interviews were undertaken with industrial practitioners with recent experience developing or migrating their legacy applications to serverless.

### 3.1 Interviews

**Interview Instruments:** the semi-structured interview guide was developed based on the research questions following the guidelines of Robinson [18]. The interview guide covers demographic information, migration intentions, strategies followed during the migration, and technical and organizational challenges. All the authors were involved in developing the interview questions by conducting regular meetings. The interview guide can be found in our replication package [19].

**Recruiting Participants:** the first and third authors attended several technology innovation industrial meetups where companies participated to share their success stories. Both authors randomly contacted industrial practitioners and asked whether they employed serverless computing in their industry. In addition, the second author contacted the targeted population by leveraging social media platforms (e.g., LinkedIn, ResearchGate, Facebook). A total of 38 participants were contacted, of which 15 were selected for the interview. We adopted a defined set of acceptance criteria for selecting our interviewees and case organizations. Mainly, our participants are (a) professionals in software engineering who (b) have either participated in or closely observed a serverless migration project within their professional scope.

We shared the interview script with the practitioners beforehand to familiarize them with the study. The interviewees were informed that ethical standards would be maintained by ensuring the confidentiality of their recordings and transcriptions. We interviewed 15 professionals from 4 countries (Finland, The Netherlands, UAE, and Pakistan) that have worked at medium to large companies in different business domains. The first author conducted all the interviews online using Zoom and Microsoft Teams platforms. The interviews lasted for ~40 to ~55 minutes on average. The recorded interviews were transcribed for further analysis. However, we omitted the interview transcripts from the replication package to ensure confidentiality [19].

### 3.2 Data Analysis

This study used a thematic analysis approach to identify, analyze, and report the findings [20]. The thematic analysis enabled us to identify motivations, strategies, and challenges subsequently mapped into themes. We utilized NVivo qualitative data analysis tool to identify and categorize the codes into themes. Initially, we meticulously read the interview transcriptions and made observational notes without establishing codes. After familiarization, we began coding the transcriptions, subsequently scrutinizing, and categorizing the resultant codes under main themes. The main themes were intentions, strategies, and technical and organizational challenges. The coding part was revisited repeatedly, and statements with similar meanings but different phrasing were connected.

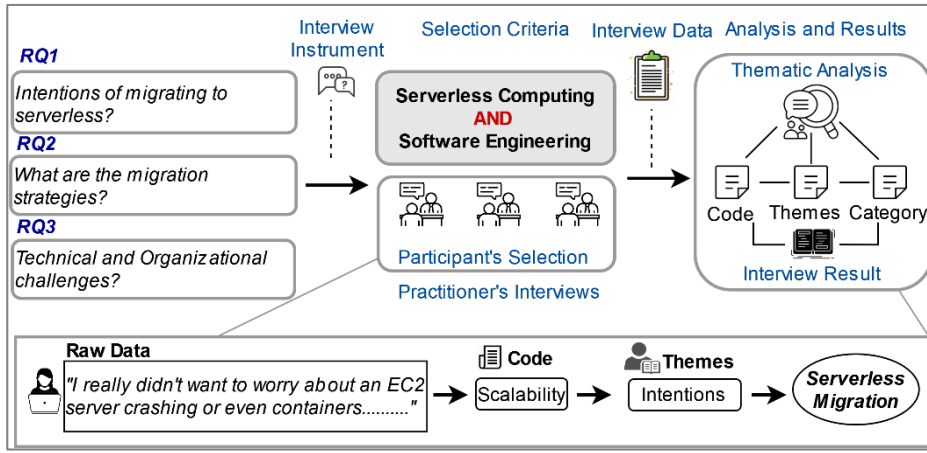

**Fig. 1.** Overview of the research methodology

## 4 Interview Case Descriptions

In this section, we describe the interview cases of systems that have been migrated to serverless or developed greenfield. To understand the systems, we only selected the participants involved in the migration process or development. All the interview participants were professional in their technical roles (e.g., lead engineer, architect, software engineer) with significant experience in the software industry and a minimum of 2 years of experience in serverless. We also interviewed participants having managerial roles in their organization to understand the organizational changes that happened due to the migration to serverless. We recruited participants from multiple countries (Finland, The Netherlands, UAE, and Pakistan) to generalize the study results. The detail of participating companies (C1-C11) of different sizes and domains are shown in Table 1. Altogether we investigated 11 systems with 15 practitioners from 11 organizations. We have presented all the migration intentions, strategies, and technical and organizational challenges reported in the interviews in our replication package [19]. However, due to the page limitation, we only present the most common intentions, strategies, and challenges in this paper. The details of the systems are discussed in the following case descriptions based on the interviews.

**C1-S1 Logistic Management System:** C1 is a large-scale enterprise providing logistic services such as mail and parcel delivery and e-commerce solutions at domestic and international levels. The has big monolithic systems to handle logistic services with physical servers. The company realized that running data centers and managing servers and operations were not aligned with its core business. It spent lots of time managing underlying infrastructure rather than focusing on business logic. Similarly, the system faced scalability, flexibility, and high-cost issues due to the high request volume, mainly in the peak season. According to P1, *"we leveraged the serverless computing as we are determined to enhance our operational efficiency, elevate customer experiences, and faster release to the market"*. The company closely analyzed the nature of the system as it was responding to several internal and external events as P1 stated *"So, a consumer might want to send a parcel and they register that parcel with one of our APIs that is an event parcel registered then we pick it up with the van and that's an event parcel picked up[…] you can think of a million other events in that system. So, personnel should build an event driven application landscape because it's the best fit to our business"*. They extracted the functionality using the bounded context of the domain-driven approach and decomposed it into more granular microservices. This allowed them to isolate and manage different functionalities independently. Following the strangler pattern approach, they gradually migrated the system to the serverless. However, they faced testing the event-driven architecture challenging due to the distributed nature of applications and the difficulty in estimating functional timeouts for long event-driven chains. Furthermore, implementing schema validation and versioning in a distributed, event-driven system, integrating with the legacy system "*Somebody has had to make a plan on where it's going and where it needs to be and which […]"* appeared as a challenge. However, they apply contract testing, unit, integration, service, and end-to-end testing depending on the need of use case. Among organizations, changing the mindset is challenging for existing developers *"Well, there's a lot of organization and a lot of cultures that you need to build around serverless"*. However, P1 stated that migrating to serverless significantly reduced their operational overhead, improved scalability, led to faster time to market and reduced cost.

**C2-S2 Financial Service Management:** the company offers a Software as a Service (SaaS) system for banks, asset managers, and other financial institutions to handle savings, current accounts, loans, mortgages, and investment products. The system can be integrated with a bank's existing systems through APIs to use its services. Initially, the system was a monolithic system hosted on AWS EC2. However, the system faced scalability issues and released new features to the market due to massive access through the API. To mitigate this, they initially understood that the nature of the system is event-driven, so the serverless architecture would be the best option in their case. They migrated their monolithic systems to monolithic Lambda using lift and shift approach as P2 stated, *"So, what we did is slowly transition to Lambda and Fargate for massive workload. […], I think it's called shift and lift. Essentially this approach embeds a whole Web API within a Lambda […]"*. The company soon realized that having a single lambda function to handle all the operation makes the system more complex to manage. However, they migrated using strangler pattern by understanding the function domain.

Furthermore, to handle substantial workloads that exceeded the service limitations of the platform, such as memory and timeout constraints, they employed a combination of Lambda functions and serverless containers (AWS Fargate) approach. The system faces several technical challenges, such as integration to an existing system, event contract validation, maintaining events order, and cold start due to the tech stack, such as .Net. Due to performance issues of other programming languages, they decided to use JavaScript and Python. They faced organizational challenges such as regulatory compliance, learning new technologies, and serverless mindset. However, migration to serverless improved scalability, faster time to market, and reduced cost.

**Table 1.** Company and participants demographics

| Comp* | Domain | Emp* | Par* | P.Role | Exp* (Ser*) | ID |
|---|---|---|---|---|---|---|
| C1 | Logistics services | 37365 | P1 | Architect | 18(5) | S1 |
| C2 | Financial Services | 17500 | P2 | Lead Engineer | 13(4) | S2 |
| C3 | E-commerce | 15000 | P3 | Architect | 16(5) | S3 |
| C4 | Web Services | 14500 | P4 | Architect | 9(3) | S4 |
| C5 | E-commerce | 9500 | P5 | Architect | 13(5) | S5 |
| C6 | E-commerce | 700 | P6 | Developer | 8(3) | S6 |
|  |  |  | P7 | Lead Engineer | 12(4.5) |  |
| C7 | Financial Services | 536 | P8 | Architect | 15(4) | S7 |
| C8 | Consultancy | 20 | P9 | Lead Engineer | 5(2) | S8 |
|  |  |  | P10 | Architect | 10(4) |  |
| C9 | AI & Security Services | 300 | P11 | Software Engineer | 5(2) | S9 |
| C10 | Financial Services | 280 | P12 | Team Manager | 5(2) | S10 |
|  |  |  | P13 | Developer | 4(3) |  |
| C11 | Media Services | 350 | P14 | Software Engineer | 6(2) | S11 |
|  |  |  | P15 | Architect | 12(3) |  |

**C3-S3 E-commerce**: the company provides e-commerce services mainly for ordering food and grocery items. Initially, the company had a big monolithic system that faced, scalability issues in the peak seasons as their traffic was unpredictable, faster time to market, and high operation overhead (e.g., managing EC2 instances). The P3 stated that *"We were determined to build something that we could own and iterate on quickly. However, there was a specific part of the system that concerned me regarding scalability. [...] EC2 server crashing or managing containers and similar backend complexities"*. They decided to identify the main functions of their system using domain driven design and gradually migrated functionalities to serverless 'strangler pattern'. *" Our strategy closely follows the 'strangler pattern,' where we begin with a low-risk component and then gradually tackle other parts of the system [...]"*. They identified the low-risk components and migrated to serverless, eventually continuing to mission-critical components upon success. The company decided to go with the serverless-first mindset for all the new applications and rewriting some of the existing tools *" is a system we designed to supplant any of these older ETL tools"*. They faced various technical challenges such as lack of standardization, platform constraints (memory or timeout), and organizational challenges like hiring serverless experts. However, migration to a serverless improved the scalability, reduced operational overhead, and cost.

**C4-S4 Pitch Decker:** the company helps other startups with various aspects, such as pitching to investors and getting up and running the startups. Initially, they used

AWS EC2 instances for hosting but encountered scalability and maintenance issues. They then switched to Elastic Beanstalk, which offered some scalability features. However, as the system grew, the scalability issue raised, so they embraced serverless architectures as P4 stated, *" that initially got us up and running, but we soon encountered issues with scalability and server maintenance, which became a significant hassle"*. The company tested components with serverless and rewrote their application. They leveraged GraphQL and GraphQL Federation to distribute the backend logic across different microservices. However, complex workflow orchestration, testing, and debugging became challenging for the organization.

**C5-S5 Online real estate system:** an intelligent home decision system that aims to help people determine the value of their homes, find their next place, and sell their current homes. Initially, the system was built monolithic but faced the scalability issues stated by P5 *"as the company grew and the volume of data it handled increased; it became clear that a more scalable, flexible solution was needed"*. Then the company explored the possibility of serverless architecture and gradually migrated to new technology. P5 stated, *"The transition to serverless was not an overnight change but a gradual process. we started by experimenting with AWS Lambda for small, non-critical tasks [...] the company began to shift more of its operations to serverless platforms"*. They face integrating the new system with its legacy one as a main pain point, *"having serverless side by side to legacy systems for example or to non-serverless systems for example is always a bit of a tricky"*. The migration improved the scalability, performance, and reduced the operational overhead *"The move not only improved scalability and performance but also reduced operational costs and complexity"*.

**C6-S6 E-commerce:** the company specializes in providing custom apparel and accessories to their customers, where they can design their own t-shirts, sweatshirts, and other items using the design tools. The company was facing the high cost of managing the servers and scalability issues as they received unpredictable workloads during and of seasonal time; stated by P6, *" we needed to transition to a more performant, scalable system, where scaling up EC2 instances was no longer a constant requirement"*. To this end, the company started exploring more scalable solutions. Initially, they did a lift and shift to cloud *" around three or four years ago, when we undertook our significant lift-and-shift migration into the cloud"* and then gradually migrating to serverless *"I've gradually been adopting and learning serverless techniques to optimize many of our more demanding workloads, particularly those under stress as we begin extracting them from larger applications"*. According to P6, *"it has always been good for us to start with monolithic serverless application and then gradually migrating to microservices serverless"*. From microservices they identified the use cases to be migrated to serverless *there's never a single approach, and I tend to favor small apps over monoliths, breaking things up only when business success indicates the need to do so"*. However, the system faces technical challenges such as testing system, event validations, integration to legacy application stated by P7: *"....and then, how did that kind of integrate back into the rest of the application"*. Similarly, changing the mindset of existing developers has seen significant organizational challenges. The application is still transitioning to be fully serverless, saving up to 90% of running application cost by improving scalability.

**C7-S7 Financial services provider:** the system provides financial services, including life and non-life insurance, retirement services, investments, and banking, to individual, corporate, and institutional clients. Their main intention was to make their system more secure, reliable, scalable, and performance *"you know, in this financial industry, it is important to take care of the data of our customers, make sure that we don't lose them, that they can always access their data when they need it. So, reliability super important performance is important"*. The system is still in migration process toward embracing fully serverless. According to P8, they tried new functions building with serverless first mindset *"We started in 2019 and started playing around with serverless with just one or two teams trying things out, but now we already have with it 20 to 25 teams actively using serverless. We are pretty much, you know, Serverless first and trying to do as much serverless as possible"*. They gradually migrated legacy systems to serverless by decomposing and testing the functionalities with serverless. Their main challenge was identifying the use case that can be migrated to serverless and enabling their existing team to adopt serverless architecture. Integrating with the existing system and evaluating the different tech stacks for better performance is also challenging. However, embracing and migrating to serverless has reduced cost, time to market, and operational overhead.

**C8-S8 Reinsurance system:** the small size company provides consultancy to other companies that tend to do greenfield development and migrate their legacy systems to serverless architecture as stated by P9: *"The projects we undertake are mostly greenfield developments, while a significant portion involves migrating large, legacy workloads to serverless platforms"*. The company recently migrated a system that provides reinsurance and insurance services, focusing on risk management and quick claim settlements. The system also creates investment opportunities aligning with their risk selection and portfolio management capabilities. The system was developed on-premises and does high performances computing problems by doing a lot of statistical modeling jobs. The system was growing fast, and it tended to improve its scalability. The existing system was taking hours to compute the business functionalities, as stated by P10: *"When we were in a position where like they were running on-prem their workload was taking quite a long time to run, but their business was growing. So, they're really talking about massive scale that are on prem servers wouldn't really handle"*. So, initially, the company started with an approach called lift and shift to migrate the workload to AWS. The company analyzed the use cases of the system and identified the domains of the system, and gradually migrated it to serverless. They faced technical challenges such as a lake of serverless job scheduling services for managing highly intensive computing, concurrency issues, debugging, and distributed logging and tracing. To mitigate these, they built several local tools. The migration to serverless architecture resulted in several benefits, including running roll-ups in under an hour, increased scalability with no limits on the number of runs, reduced code base by about 70%, improved cost efficiency, and decreased operational overhead.

**C9-S9 Smart mobility system:** the startup company developed the smart mobility data generation system. This system involves collecting data from mobile phones and sending it to the startup's backend infrastructure. The startup wants to develop a system where they reduce the cost of the system and manage the underlying infrastructure, as stated by P11: *"The need for scalability and flexibility in their operations was paramount. We want to get rid of like the time we spent on managing servers"*. The company evaluated that the nature of its system is event-driven and will grow exponentially, so it decided to go with a serverless first mindset. They initially developed the monolithic serverless system. As the monolithic system reached complexity, they migrated to microservices serverless to reduce the system's complexity. They decomposed the system using a domain-driven design approach and delegating the Lambda microservices by teams. However, the startup faces the testing and debugging of an event-driven application challenging. The application solved the scalability and cost issues by embracing serverless.

**C10-S10 Financial services management:** the system provides bookkeeping, invoicing, and banking functions, enabling small and medium-sized businesses to streamline their financial operations and improve overall business functions. Initially, they had a big monolithic system that handled all the operations. However, over time, the system started facing scalability issues and reduced time to market for new features. The system is still in transition from monolithic to microservices. They initially evaluated the uses cases that were best suitable for serverless and rewrote the entire use case as stated by P12: *"We are basically rewriting some parts of our system. We are of course trying not to directly, you know, cut out pieces, but moreover rethinking of serverless"*. The new use cases are then integrated with the system through API. The company rethinks the purpose of the new features and works on them separately to ensure smooth integration with the current infrastructure.

Furthermore, they utilized the Fargate and Lambda as a combination for massive workloads. Testing and integrating with existing systems are noted as significant challenges. The adoption of serverless reduced operational overhead, improved scalability, and the ability to focus on business logic rather than infrastructure management.

**C11-S11 Digital media services:** The company provides several media services, including television, radio, and digital media. Among many applications, the digital media services system distributes formats and content to other media companies and social media platforms. Initially, the system was developed using a monolithic approach on EC2 machines. Sooner, the system started facing scalability issues as the number of users increased stated by P14: *" The problem at that time was that the team was constantly firefighting the system. It couldn't scale, and every time we adjusted something, it would break something else in the distributed monolithic world"*. They started evaluating the use cases and ended up with serverless. The company gradually migrate to serverless using strangler pattern P15: *" We employ the strangler pattern, incrementally replacing components of the legacy system with a serverless model, aiming to complete the transition within a six-month timeframe."*. Serverless architecture greatly helped them by simplifying rules, designing functions with single responsibilities, and improving flow with Step Functions. The main challenge was to migrate the data from the existing monolithic database to microservices. Instead of rewriting the entire system, they updated the legacy system to emit events; when some change happens to the database, it emits the event and stores the events in the new database. This way, they migrated the database. They gradually created parallel microservices to handle specific functionalities previously performed by the monolith. Over time, they redirected requests from the monolith to the new microservices, repeating the process gradually. However, migrating to serverless reduced the cost significantly. The team experienced significant speed improvements, delivering features in weeks rather than months, with reduced bugs and improved testing practices.

## 5      Results and Discussion

This section describes the aggregated results of our study. All the results of our research questions are presented in Fig.2. We only presented the common findings in more than two systems. However, all the findings from the participants and migration cases are presented in our replication package [19].

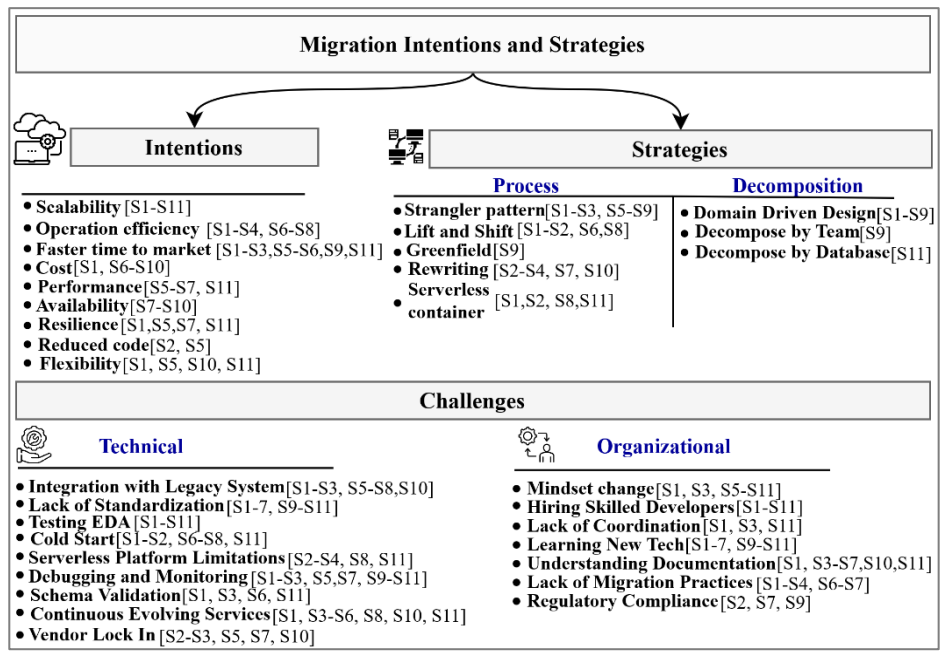

**Fig. 2.** Migration intentions, strategies, and challenges

### 5.1      Migration intentions (RQ1)

The Fig.2. presents the intention to migrate systems, strategies, and challenges (technical and organizational) to a serverless architecture.

**Drivers for migrating legacy systems:** our study reveals that large-scale enterprise systems face scalability issues. Participants (P3, P6, P8, P9, P11, P12, P13) stated that their systems were receiving unpredictable traffics, mainly on seasonal days, causing delays in response. They spent most of the time provisioning the underlying infrastructure rather than focusing on the business logic. This leads them to migrate to serverless architecture *" We were determined to build something that we could own and iterate on quickly. However, there was a specific part of the system that concerned me regarding scalability. I didn't want to have to worry about issues like an EC2 server crashing or managing containers and similar backend complexities"*. Apart from these, the companies were facing issues such as a lack of system overview (P1), downtime during updates (P13), and costly changes (P15).

Similarly, time to market was the main driver for most of the large-scale organizations. New features take over 6 months for several organizations to go into production. However, for small-scale organizations, the main driver for adopting serverless is to reduce the cost and operational overhead. Participant P4 stated: *"I had my own startup where every dollar counted"*. Apart from these main drivers of migrating the legacy system to serverless, (P3) resilience, (P5) reliability, (P2) flexibility, (P6, P7) performance, (P9, P10) reduced code base and increased agility, (P12, P13) maintainability are considered the main drivers for migrating legacy system to serverless architecture.

### 5.2   Migration Strategies and Decomposition (RQ2)

The prevalent migration strategy in our investigated systems was identifying the use case that had the best fit to the serverless. Seven participants identified that their systems have an event-driven nature and has unpredictable workloads. They extracted the use cases from the existing system, built and tested it with a serverless system, and integrated it with the legacy system. This process leads them to gradually migrate their entire system to serverless using strangler pattern as in P5: *"We started in 2019 and started playing around with serverless with just one or two teams trying things out, but now we already have with it 20 to 25 teams actively using serverless"*. All the interviewed participants stated that they built new functionality with the serverless first mindset and integrated it with the legacy system. They were more focused on developing new projects with the serverless first mind set.

Furthermore, we investigated two high-computing systems doing a lot of statistical computation. The systems were not the best fit for serverless due to the resource limitations of serverless platforms. Thus, they used a combination of serverless and containers to compute the massive workload *"So, we used Fargate for long-running massive workloads due to its improved scalability characteristics. On the other hands, Lambda was used for near real-time analytics where quick response times was crucial"*. Five participants were involved in rewriting the existing functionality with serverless architecture.

Similarly, four participants said they made a lift and shift approach from on-premises to cloud serverless. One system was developed with a greenfield development approach where the participant predicted that their system would receive a high workload in the coming years, and they wanted to reduce the cost and time of managing the underlying infrastructure. Companies start with monolithic lambda functions for greenfield projects to offer the application's functionalities. However, the complexity of managing monolithic functions increases as the function grows. Therefore, the best practice is to identify and split the monolithic serverless to serverless microservices, as reported by participants. Most of the systems were migrated in two years, and new functionality was released in hours.

For the migrated system to serverless architecture, identifying the use case and boundaries of the services was challenging for many companies. Nine out of eleven companies adopted the domain-driven design approach for identifying the boundaries and splitting it into Lambda microservices. Two companies decomposed their systems based on database complexity or API integration. One system decomposed the system by teams. The participant revealed that there are no automatic tools to identify the use cases and boundaries from existing systems.

### 5.3 Challenges (RQ3)

The third section of the figure presents the technical and organizational challenges companies face when migrating their legacy systems to a serverless architecture.

**Technical challenges:** the migration journey poses several technical challenges. Identifying the use cases best fit for the serverless architecture and splitting the system was challenging for organizations. Furthermore, integration with legacy systems is reported as a significant challenge. P5 stated, *"If you want Greenfield, then take this toolbox and go ahead. But now we also have lots of like legacy bits in our system [……] to integrate all these systems is a technical challenge"*. Similarly, lack of standardization, testing event-driven architecture, validating the events, building the central event broker, and managing different accounts are also prevalent challenges.

Furthermore, financial systems are more concerned with vendor lock-in problems due to regulatory compliance. Companies utilizing the old tech stack are more prone to face performance issues due to the cold start problem of serverless. Programming languages like Java, Python, and Typescript hold fewer cold start problem than other languages such as C# and .Net. Therefore, choosing the right tech stack is becoming challenging for organizations that are using old programming languages, i.e., C#, .Net. Moreover, handling the long-running process becomes a challenge due to serverless platforms' resource and time limitations. Most organizations built a charter to standardize the use of technology stack. They standardize around dev, test, and observability *"We standardized around a dev, test, acceptance, and production workflow and currently, we're looking at standardizing around observability tooling"*.

**Organizational challenges:** our study revealed that organizations face several changes when migrating their legacy systems to serverless. The transition from traditional to more agile development needs a change in the organization's mindset. Convincing developers to develop the application serverless is sometimes challenging as they are stuck to the traditional way of development, as stated by P4: *"You know, this served as the first mindset, partly because it's really difficult to get buying from so many developers who are used to building things traditionally"*. To convince more developers to adopt serverless, the serverless evangelist started testing new components with a few team members and presented the success within other teams of the project *"first year, we were tasked with doing all of these corporate startup projects and we did something around 10 to 20 projects with serverless first mindset, we started in a very small corner of our company in innovation and then we presented the success with*

*others"*. Some organizations hire serverless evangelists and develop a serverless implementation guide to follow the process. Learning new technology and enabling ops mindset in the development team for serverless is also challenging for many organizations *"So really bringing the DevOps mentality that you build it, you run it in team is challenging"*. For this, organizations encourage attending serverless workshops, taking courses, and participating in serverless success stories. Furthermore, hiring skilled serverless developer (P1-P13), lack of understanding the platform documentations (P3, P6, P7, P9, P13), lack of existing migration practices (P1, P3, P6, P7, P9, P15) and lack of coordination (P2, P4, P11) also seen as a challenge in the organizations.

## 6  Threat to Validity

Several potential threats could impact the validity of the results of this study. These threats are typically categorized into four primary categories: internal validity, construct validity, external validity, and conclusion validity [21].

**Internal validity:** refers to the degree to which specific factors influence methodological robustness. The first threat to this study is the understanding of the interview questions by participants. We mitigated this threat by conducting the pilot interviews with other professionals of our contacts and by sharing the questions beforehand to ensure the interview questions' understandability and readability. Only participants having knowledge of the migration process or serverless greenfield projects are included in the final interviews.

**Construct validity:** refers to the degree to which the research constructs are adequately substantiated and interpreted. The core constructs are the interview participants' viewpoints on the migration or adoption of serverless technology and the associated challenges in the context of this study. The verifiability of the construct is considered the limitation of qualitative studies. Therefore, we followed a rigorous and step-by-step research method process and gave examples in quotations from the collected data (e.g., interviews). This shows how a well-planned research process and the findings we reported (RQ1, RQ2, RQ3) help prove the verifiability of the study construct.

**External validity refers to generalizing** the conducted study to another context. The sample size and sampling approach in the study may not allow generalizing the findings. Also, serverless architecture and migration of the legacy applications to serverless is not yet well established in practice. Therefore, finding a large sample was challenging to us. However, we mitigated this threat by using all possible sources to find the potential population. We collected data from 4 countries across two continents from participants with diverse experience in various industrial domains. In the future, we plan to extend our study findings by mining the Q&A platforms and conducting industrial surveys and interviews.

**Conclusion Validity:** refers to the factors that impact the trustworthiness of the study conclusion. To mitigate this threat, all the authors conducted weekly meetings to develop the interview instruments and data analysis process. The first author conducted the interviews via Zoom and Microsoft Teams. All other authors reviewed the data and provided feedback to improve the analysis. Finally, all the authors conducted a brainstorming session to draw the findings and conclusion of this study.

## 7　Conclusion and Future Work

In this study, we investigated 11 systems from 11 organizations by conducting interviews with 15 participants having diverse professional experiences in various domains. We mainly investigated intentions, strategies, and challenges (technical and organizational) during migrating legacy applications to serverless. The results revealed scalability was the main challenge for large-scale enterprises due to unpredictable traffic. They were spending much of the time managing underlying infrastructure rather than focusing on the business logic, which hindered them from releasing features faster in production. On the other hand, smaller organizations tended to reduce the cost and operational overhead.

The study revealed that most systems were migrated gradually using strangler patterns to serverless, while new components were developed with a serverless first mindset. However, legacy monolithic systems were decomposed into microservices and migrated to serverless by applying domain driven design approach, decomposition by database, and decomposition by teams. Integration into the existing system, testing event-driven architecture, and validating the event schema were reported as the main technical challenge. The organizations face severe mindset changes of existing developers toward the new technology, hiring the serverless evangelist, and lack of existing migration practices as the main organizational challenge during migration.

In future work, we plan to expand our research to various industry domains to uncover more practical challenges and solutions related to serverless migration. We also aspire to holistically chart the decision-making process on all the dimensions of migration to serverless. Furthermore, we will develop a multicriteria decision-making (MCDM) framework for platform selection in the industry.